\documentclass[12pt,a4]{article}
\begin{document}
\setlength{\textheight}{9.4in}
\setlength{\topmargin}{-0.6in}
\setlength{\oddsidemargin}{0.2in}
\setlength{\evensidemargin}{0.5in}
\renewcommand{\thefootnote}{\fnsymbol{footnote}}
\begin{center}
{\Large {\bf Supersymmetry in Quantum Mechanics of Colored Particles}}
\vspace{0.8cm}

{Sh. Mamedov,\raisebox{0.8ex}{\small a,b}\footnote[1]
{ Email:shahin@physik.uni-kl.de}
Jian--zu Zhang\raisebox{0.8ex}{\small a,c}\footnote[2]
{Email:jzzhang@physik.uni-kl.de}
and V. Zhukovskii\raisebox{0.8ex}{\small d}\footnote[3]
{Email:th180@phys.msu.su}\\

\raisebox{0.8 ex}{\small a)}{\it Department of Physics, University
of Kaiserslautern, 67653 Kaiserslautern, Germany}

\raisebox{0.8 ex}{\small b)}{\it  High Energy Physics Laboratory, Baku State
University, 23 Z. Khalilov, 370148 Baku, Azerbaijan}

\raisebox{0.8 ex}{\small c)}{\it Institute for Theoretical Physics, Box 316,
East China University of Science and Technology, Shanghai 200237, P.R. China}

\raisebox{0.8 ex}{\small d)}{\it Faculty of Physics,
Department of Theoretical Physics, Moscow
State University, 119899 Moscow, Russia}}
\end{center}
\vspace{0.3cm}

{\centerline {\bf Abstract}}
\noindent

The role of supercharge operators is studied in the case
of a Dirac particle moving in a constant
chromomagnetic field. The Hamiltonian is factorised and the ground state
wave function in the case of unbroken supersymmetry
is determined.

\vspace{2cm}
The study of supersymmetric quantum mechanics began with
 ref.\cite{1} and has been  applied to
various problems, including in particular the case of
charged abelian particles\cite{2,3,4,5,6,7}.
In refs.\cite{2} and \cite{3} the supersymmetry of
the equation of motion of the electron in
a magnetic field was investigated. The related
problem of supersymmetry of the Dirac equation
for a colored particle in an external
chromonagnetic field has been considered in
ref.\cite{8}. In the following the investigation of the
latter is continued with the aim to explore the significance of
the supercharge operators in this case of
a chromomagnetic field, to factorise the Hamiltonian
and to determine the ground state
wave function.

It is known \cite{9,10} that a constant color background
can be obtained from two types of vector potentials
$A_{\mu}$, i.e. linearly growing abelian potentials and non-commuting
vector potentials.  Here we consider the
simplest case of the latter type which implies
the non-commuting potentials
\begin{eqnarray}
A_{\mu}&=&A^a_{\mu}T^a=A^a_{\mu}\tau^a/2, 
\;\;a=1,2,3, \;\; \mu = 0,1,2,3,\nonumber\\
 A^1_{\mu}&=&(0,\sqrt{\lambda_1},0,0), A^2_{\mu}=(0,0,\sqrt{\lambda_2},0),
A^3_{\mu}=0.
\label{1}
\end{eqnarray}
with $\sqrt{\lambda_1},\sqrt{\lambda_2}$ constant.
Latin indices are $SU(2)$ color indices and Greek indices
are Lorentz ones. The Pauli matrices $\tau_i$ are the generators of the
$SU(2)$ color group. The potentials (\ref{1})
determine a constant homogeneous chromomagnetic field
along the third axes of both ordinary and color spaces, 
i.e.
\begin{equation}
F^3_{12} = g\epsilon^{ab3}A^a_1A^b_2 = g\sqrt{\lambda_1
\lambda_2} = H^3_z
\label{2}
\end{equation}
where $g$ is the constant of color interaction.  
With analogous non-commuting
potentials a constant field could be
given for any chromomagnetic or chromoelectric
component.  

The Dirac equation for a colored
particle in the
chromomagnetic field (\ref{1}) is 
\begin{equation}
\bigg[\gamma_{\mu}\bigg(p_{\mu}+gA^a_{\mu}\frac{{\tau}^a}
{2}\bigg)-m\bigg]\psi=0.
\label{3}
\end{equation}
In terms of
Majorana spinors $\phi$ and $\chi$ $\psi =\left(
\begin{array}{c}\phi\\\chi \end{array}\right)$ and
with
$P_{\mu}=p_{\mu}+gA^a_{\mu}{\tau}^a/2$
we rewrite eq.(\ref{3}) as
\begin{eqnarray}
\sigma_i P_i\chi
 &=& \bigg(i\frac{\partial}{\partial t}-m
\bigg)\phi,\nonumber\\
\sigma_i P_i\phi &=& 
\bigg(i\frac{\partial}{\partial t}+m
\bigg)\chi
\label{4}
\end{eqnarray}
Here the Pauli matrices $ \sigma_i$ describe the particle's spin.
The spinors $\phi$ and $\chi$
can easily be seen to satisfy the equations
\begin{eqnarray}
\bigg({\bf \sigma}\cdot{\bf P}\bigg)^2\phi &=&
-\bigg(\frac{\partial^2}{\partial t^2}+m^2\bigg)\phi,\nonumber\\
\bigg({\bf \sigma}\cdot{\bf P}\bigg)^2\chi &=&
-\bigg(\frac{\partial^2}{\partial t^2}+m^2\bigg)\chi
\label{5}
\end{eqnarray}
Setting
\begin {equation}
Q={\bf \sigma}\cdot{\bf P}
\label{6}
\end{equation}
and
\begin{equation}
H=-\bigg(\frac{\partial^2}{\partial t^2} + m^2\bigg)
\label{7}
\end{equation}
eqs. (\ref{5}) assume the form
\begin{equation}
Q^2\psi = H\psi.
\label{8}
\end{equation}
The ``Hamiltonian'' (\ref{7}) (cf. ref.\cite{2})
 has eigenvalues $E^2-m^2$, i.e.
\begin{equation}
H\psi = (E^2-m^2)\psi.
\label{9}
\end{equation}
Eqs. (\ref{8}) are reminiscent of Witten's one--measure
supersymmetric quantum mechanics \cite{1}
for $N$ supercharges. In
this the supersymmetry algebra
has the form
\begin{equation}
\{Q_i, Q_j\}= 2\delta_{ij}H, \;\;\;
 [H, Q_i]=0, \;\;\; i=1, 2, 3, \cdot\cdot\cdot N
\label{10}
\end{equation}
where the curly bracket denotes an anticommutator.
In view of its commutation
with $H$ the quantity ${\bf \sigma}\cdot{\bf P}$ is conserved on
the classical level.
It is known in supersymmetric theories that supercharge operators
$Q_i$ obeying (\ref{10}) lead to a degeneracy of the energy.
The relation between the
degeneracy $n$ and the number of supercharges $N$
is given by the formula
\begin{equation}
n=2^{[N/2]} 
\label{11}
\end{equation}
where $[N/2]$ means {\it integer part of} $N/2$.  
The energy spectrum of a colored particle in the field
(\ref{1}) was  found in \cite{10} and is given by
\begin{equation}
E^2_{1,2}={\bf p}^2 + m^2 +\frac{g^2(\lambda_1+\lambda_2)}{4} 
\pm g\sqrt{\lambda_1 p^2_1+\lambda_2p^2_2+\frac{(H^3_z)^2}{4}}
\label{12}
\end{equation}
This spectrum is doubly degenerate so that two
energy levels correspond to four states
having different quantum numbers of $(\sigma, \tau)$.

From the explicit expression of the operator $Q$  of eq.(\ref{6})
it is seen that supersymmetry transformations
transform a state with quantum numbers
$(p, \sigma, \tau)$ into another state with
quantum numbers $(p^{\prime\prime}, \sigma^{\prime\prime},
 \tau^{\prime\prime})$. There are states with the same energy among these.
The states with the same energy
are called superpartners.
There are therefore superpartner states on the field of
eq.(\ref{1}).
According to eq.(\ref{11}) there is a second
supercharge operator in the theory.
Since $A^a_3=0$ the z--dependence of the quark wave
function is of the plane wave type. Considering
the case $p_z = 0$ at initial time $t=0$, 
 it is reasonable to consider the supercharge $Q$
in the $(x, y)$ plane and set
\begin{equation}
Q_1=\sigma_1P_1 + \sigma_2 P_2
\label{13}
\end{equation}
The second supercharge operator can be constructed with the
help of the prescription suggested in ref.\cite{2}:
\begin{equation}
Q_2=iQ_1\sigma_3=\sigma_2P_1-\sigma_1P_2
\label{14}
\end{equation}
The supercharge $Q_2$ is hermitian and
together with $Q_1$ obeys Witten's
supercharge algebra eq. (\ref{10}).
With the help of $Q_1$ and $Q_2$ we can construct 
hermitian mutually conjugate supercharge
operators $Q_{\pm}$ by setting
\begin{equation}
Q_{\pm}=\frac{1}{2}(Q_1\pm iQ_2)=Q_1\frac{1\mp \sigma_3}{2}.
\label{15}
\end{equation}
These operators obey the field theory supersymmetry
algebra\cite{11} and the Jacobi identity:
\begin{eqnarray}
Q^2_+ &=& Q^2_- = 0, \;\;\; \{Q_+, Q_-\}=H,\;\;\; [Q_{\pm}, H] = 0,\nonumber\\
0&\equiv &\{[Q_+,H], Q_-\} +\{[Q_-,H], Q_+\} +[H, \{Q_+,Q_-\}]
\label{16}
\end{eqnarray}
Setting
$$
P_{\pm} = P_1\pm iP_2, \;\;\;a_{\pm}=\frac{1}{2}(\sigma_1 \pm i\sigma_2),
$$
which satisfy $P^{\dagger}_{\pm}=P_{\mp}, a^{\dagger}_{\pm}=a_{\mp}$, 
the operators $Q_{\pm}$ can be expressed as
\begin{equation}
Q_+=P_-a_+, \;\;\; Q_-=P_+a_-,
\label{17}
\end{equation}
and the supersymmetry algebra (\ref{15}) is
fulfilled by the relations
\begin{equation}
[Q_+, P_-]=0, \;\;[Q_-, P_+]=0, \;\;\{Q_+, a_-\}=P_-,
\;\;\{Q_-, a_+\}=-P_+,
\label{18}
\end{equation}
which is in full agreement with ref.\cite{11}.
Only four of the six operators are independent.

It is easy to see from the definition of the operators $a_{\pm}$
that they obey the algebra of creation and annihilation operators
for fermion degrees of freedom, i.e.
$$
\{a_+,a_-\}=1, \;\;\; (a_+)^2 = (a_-)^2 = 0.
$$
This implies that the fermion degree of freedom in the
quantum mechanics
of the nonabelian charged particle in a chromomagnetic field is
the particle's spin. The commutator of the operators
$P_{\pm}$ gives
\begin{equation}
[P_+, P_-]=gH^3_z\tau_3
\label{19}
\end{equation}
 The spinors $\phi$
and $\chi$ of eqs. (\ref{4})  and (\ref{5})
 have two components corresponding to different
projections of spin which we write $\psi =\left(\begin{array}{c} \psi_+\\
\psi_-\end{array}\right)$.
 The components $\psi_+$ and $ \psi_-$
are spinors in
color space which  we write  
$\psi_+ =\left(\begin{array}{c} \psi^{(1)}_+\\
\psi^{(2)}_+\end{array}\right)
, \psi_- =\left(\begin{array}{c} \psi^{(1)}_-\\
\psi^{(2)}_-\end{array}\right)$.
Since the color operator $\tau_3$ has its own two eigenvalues, the commutator
(\ref{19}) splits into two commutation relations
depending on the color state chosen, i.e.
\begin{eqnarray}
&a)&\;\; [P_+,P_-]=gH^3_z \;\;\;
 for \; state \; \psi^{(1)}_{\pm} \; with \;\;\;
 {\hat \tau}_3\psi^{(1)}_{\pm}=\psi^{(1)}_{\pm},
\nonumber\\
&b)& \;\;[P_+,P_-]=-gH^3_z \;\;\; for \; state 
\; \psi^{(2)}_{\pm} \;  with \;\;\;
  {\hat \tau}_3\psi^{(2)}_{\pm}=-\psi^{(2)}_{\pm}
\label{20}
\end{eqnarray}
We run into a situation which is analogous to one
in ref.\cite{12} where the quantum mechanical motion of
a wave packet was shown to be made up of a mixture of
states with $\tau_3$--eigenvalues = +1 and -1. 
In a chromomagnetic field  such
a packet breaks up into modes with different values
of $\tau_3$ moving in opposite directions.
We introduce operators $b_{\pm}$ as
creation and annihilation operators
of bosonic states respectively by defining 
$$
b_{\pm} = \frac{P_{\pm}}{\sqrt{gH^3_z}}
$$
which obey
correspondingly  the Heisenberg--Weyl algebra
\begin{equation}
a) \;\;\; [b_+,b_-] = 1 \;\; for \; \psi^{(1)}_{\pm},
 \;\;\;\; b)\;\;\; [b_-, b_+]= 1
\;\; for \; \psi^{(2)}_{\pm}.
\label{21}
\end{equation}
>From this we see, that the operators $b_+$
and $b_-$ interchange their roles
for  state $\psi^{(1)}$
and  state $\psi^{(2)}$.
The meaning of the operators $Q_{\pm}$
of eqs.(\ref{16}) becomes
clear \footnote{Here the operators
$a_{\pm}$ do not have the same meaning as in field
theory where the fermion operators
imply changes of spin by $\pm 1/2$.}:
If $n_f$ and $n_b$ denote the number of fermions
 and
bosons respectively,
the operator $Q_+$ transforms a state $\psi^{(1)}$
with $(n_f, n_b)$ into the state
$\psi^{(2)}$ with $(n_f+1, n_b-1)$
and the state $\psi^{(2)}$ with
$(n_f+1, n_b-1)$ into the state
$\psi^{(1)}$ with $(n_f, n_b)$, and
the operator $Q_-$ transforms the state
$\psi^{(2)}$ with $(n_f+1, n_b-1)$ into the state $\psi^{(1)}$
with $(n_f, n_b)$ and the state $\psi^{(1)}$ with
$(n_f, n_b)$ into the state $\psi^{(2)}$
with $(n_f+1, n_b-1)$.  In any case the sum
of fermion and boson numbers is conserved, $n_f+n_b=const.$

We now rewrite the anticommutator (\ref{16}) using (\ref{17}):
\begin{equation}
H=Q_+Q_-+Q_-Q_+=\frac{1}{2}\bigg\{P_-,P_+\bigg\}
+\frac{1}{2}\bigg[P_-,P_+\bigg]\sigma_3
=\left(\begin{array}{cc}P_-P_+&0\\
0 & P_+P_-\end{array}\right)
\label{22}
\end{equation}
Thus the Hamiltonian of (\ref{8}) is split into two
parts corresponding to different projections of spin.
According to eq. (\ref{9}) we have
\begin{equation}
H\psi=\left(\begin{array}{cc}P_-P_+ & 0\\
0 & P_+P_-\end{array}\right)\psi = (E^2-m^2)\left(\begin{array}{c}\psi_+ \\
\psi_-\end{array}\right)
\label{23}
\end{equation}
Thus we have two independent equations:
\begin{eqnarray}
P_-P_+\psi_+&=&(E^2-m^2)\psi_+,\nonumber\\
P_+P_-\psi_-&=&(E^2-m^2)\psi_-
\label{24}
\end{eqnarray}
These equations could be used for finding the ground state wave functions
of quarks in the field (\ref{1}).
It is well known that with unbroken supersymmetry
the ground state has zero energy eigenvalue. As we see from
(\ref{12}) the energy
spectrum $E^2_1$ has its minimal eigenvalue
at ${\bf p}=0$,
\begin{equation}
\bigg(E^2_1\bigg)_{min}=\bigg( m^2+\frac{g^2({\lambda_1}^{1/2}+{\lambda _2}
^{1/2})^2}{4}\bigg),
\label{25}
\end{equation}
and 
 the spectrum $E^2_2$ has its minimal eigenvalue at
${\bf p} = (0, g\sqrt{(\lambda_2-\lambda_1)}/2, 0)$
for $\lambda_2 > \lambda_1$ 
and at ${\bf p} = (g\sqrt{(\lambda_1-\lambda_2)}/2, 0, 0)$
for $\lambda_1> \lambda_2$ with
\begin{equation}
 \bigg(E^2_2\bigg)_{min} = m^2.
\label{26}
\end{equation}
Hence 
only for the spectrum $E^2_2$ supersymmetry is
unbroken for massless quarks ($m=0$).
Then for the ground state with unbroken supersymmetry
eqs.(\ref{24}) assume the form
\begin{equation}
P_-P_+\psi_+ = 0, \;\;\; P_+P_-\psi_- = 0. 
\label{27}
\end{equation}
These equations show that the Hamiltonian for the
ground state is reduced to a product of two
linear, i.e. factorised, operators.
According to (\ref{27}) the expectation values of 
the operators $P_{\mp}P_{\pm}$ vanish, i.e.
\begin{equation}
<\psi_+|P_-P_+|\psi_+> = 0, \;\;
<\psi_-|P_+P_-|\psi_-> = 0.
\label{28}
\end{equation}
Since $P_{\pm}$ are mutually hermitian  conjugate operators
these equations can be written as 
\begin{equation}
\bigg|P_+|\psi_+>\bigg|^2=0, \;\;
\bigg|P_-|\psi_->\bigg|^2=0.
\label{29}
\end{equation}
Thus to find the ground state wave function we have to solve
the two linear equations obtained from (\ref{29}). Taking
into account the explicit expressions of $P_{\pm}$ and
the color components $\psi^{(1),(2)}_{\pm}$ of the spinors
$\psi_{\pm}$, we obtain
\begin{eqnarray} 
P_+|\psi_+> &=&\left(\begin{array}{cc}
p_1+ip_2 & \frac{ig}{2}(A^1_1+A^2_2)\\
\frac{ig}{2}(A^1_1-A^2_2) & p_1+ip_2
\end{array}\right)\left(\begin{array}{c} \psi^{(1)}_+\\
\psi^{(2)}_+\end{array}\right) = 0, \nonumber\\
P_-|\psi_+> &=&\left(\begin{array}{cc}
p_1-ip_2 & \frac{ig}{2}(A^1_1-A^2_2)\\
\frac{ig}{2}(A^1_1+A^2_2) & p_1-ip_2
\end{array}\right)\left(\begin{array}{c} \psi^{(1)}_-\\
\psi^{(2)}_-\end{array}\right) = 0.
\label{30}
\end{eqnarray}
These equations are easily solved in polar coordinates
with $ x=r\cos\theta, y= r\sin\theta$, in view of the cylindrical
symmetry of the external field (\ref{1}).
In these coordinates the operators $p_1\pm ip_2$
assume the form
\begin{equation}
p_1\pm ip_2 = e^{\pm i\theta}\bigg(\frac{\partial}{\partial r}\mp\frac{i}
{r}\frac{\partial}{\partial\theta}\bigg)
\label{31}
\end{equation}
We thus have two independent variables $r, \theta$
and one constraint resulting from conservation of
$Q={\bf \sigma}\cdot{\bf P}$. 
We choose the reference frame so that $\theta$ is the
angle between ${\bf \sigma} $ and ${\bf P}$, and
we assume that the external chromomagnetic field given
by potentials (\ref{1}) is given in this
reference frame, since otherwise, 
on passing to any other moving frame, the external
field (\ref{1}) will also possess chromoelectric
components.  Conservation of ${\bf \sigma}\cdot{\bf P}$
means $\cos\theta = const.$ Consequently
$\theta = const.$ and $\partial\psi^{(1),(2)}_{\pm}/\partial\theta =0$.
With this eqs.(\ref{30}) assume the well known form
\begin{equation}
\bigg(\frac{\partial^2}{\partial r^2}+\frac{g^2}{4}e^{\mp2i\theta}
(\lambda_2-\lambda_1)\bigg)\psi^{(1)}_{\pm} =0, \;\;
\psi^{(2)}_{\pm}=\frac{2e^{\pm i\theta}}{g(\sqrt{\lambda_1}+\sqrt{\lambda_2})}
\frac{\partial}{\partial r}\psi^{(1)}_{\pm}.
\label{32}
\end{equation}
Setting
$$
\xi=\frac{g}{2}\sqrt{\lambda_2-\lambda_1}\sin\theta, \;\;
\eta=\frac{g}{2}\sqrt{\lambda_2-\lambda_1}\cos\theta,
$$
the general solution of eqs. (\ref{32}) can be written
\begin{eqnarray}
\psi^{(1)}_+&=& C_1e^{\xi r}e^{i\eta r}+C_2e^{-\xi r}e^{-i\eta r},\nonumber\\
\psi^{(2)}_+&=&i\sqrt{\frac{\sqrt{\lambda_2}-\sqrt{\lambda_1}}
{\sqrt{\lambda_1}+\sqrt{\lambda_2}}}
\bigg( C_1e^{\xi r}e^{i\eta r}-C_2e^{-\xi r}e^{-i\eta r}\bigg),\nonumber\\
\psi^{(1)}_-&=&i\sqrt{\frac{\sqrt{\lambda_2}-\sqrt{\lambda_1}}
{\sqrt{\lambda_1}+\sqrt{\lambda_2}}}
\bigg( C^{\prime}_1e^{-\xi r}e^{i\eta r}
-C^{\prime}_2e^{\xi r}e^{-i\eta r}\bigg),\nonumber\\
\psi^{(2)}_-&=& C^{\prime}_1e^{-\xi r}
e^{i\eta r}+C^{\prime}_2e^{\xi r}e^{-i\eta r}.
\label{33}
\end{eqnarray}
Selection of the normalizable parts depends on
whether $\lambda_2>$ or $<\lambda_1$ and on the value of
the angle $\theta$. For instance when $\lambda_2>\lambda_1$
and $0< \theta\leq \pi/2$, we have
\begin{eqnarray}
\psi^{(1)}_+&=&C_2e^{-\xi r}e^{-i\eta r},\nonumber\\
\psi^{(2)}_+&=&-iC_2\sqrt{\frac{\sqrt{\lambda_2}-\sqrt{\lambda_1}}
{\sqrt{\lambda_1}+\sqrt{\lambda_2}}}
 e^{-\xi r}e^{-i\eta r},\nonumber\\
\psi^{(1)}_-&=&iC^{\prime}_1\sqrt{\frac{\sqrt{\lambda_2}-\sqrt{\lambda_1}}
{\sqrt{\lambda_1}+\sqrt{\lambda_2}}}
 e^{-\xi r}e^{i\eta r},\nonumber\\
\psi^{(2)}_-&=& C^{\prime}_1e^{-\xi r}
e^{i\eta r}.
\label{34}
\end{eqnarray}

Further analysis of the Dirac equation shows that there is no
supersymmetry in the chromoelectric case. Also supersymmetry is
broken in a chromomagnetic field with spherically
symmetric components.

\vspace{0.4cm}

\noindent
{\bf Acknowledgments}

Sh. M. and J.--z. Z. acknowledge
 support by DAAD 
and discussions with H. J. W. M\"uller--Kirsten. JZZ's work has also been 
supported by the National Natural Science 
Foundation of China under the grant number 10074014 and by the Shanghai 
Education Development Foundation.


\end{document}